# Surface polar states and pyroelectricity in ferroelastics induced by flexo-roto field

A.N. Morozovska[1,2], E.A. Eliseev[2], S.V. Kalinin[3], Long Qing Chen[4] and Venkatraman Gopalan[4]

[1] Institute of Semiconductor Physics, National Academy of Science of Ukraine,
41, pr. Nauki, 03028 Kiev, Ukraine

[2] Institute for Problems of Materials Science, National Academy of Science of Ukraine,
3, Krjijanovskogo, 03142 Kiev, Ukraine

[3] Center for Nanophase Materials Science, Oak Ridge National Laboratory,
Oak Ridge, TN, 37831

[4]Department of Materials Science and Engineering, Pennsylvania State University,
University Park, Pennsylvania 16802, USA

**Abstract**
Theoretical analysis based on the Landau-Ginzburg-Devonshire (LGD) theory is used to show that the joint action of flexoelectric effect and rotostriction leads to a large spontaneous in-plane polarization ($\sim$ 1-5 $\mu C/cm^2$) and pyroelectric coefficient ($\sim 10^{-3}$ $C/m^2 K$) in the vicinity of surfaces of otherwise non-ferroelectric ferroelastics, such as $SrTiO_3$, with static octahedral rotations. The origin of the improper polarization and pyroelectricity is an electric field we name *flexo-roto field* whose strength is proportional to the convolution of the flexoelectric and rotostriction tensors with octahedral tilts and their gradients. Flexo-roto field should exist at surfaces and interfaces in all structures with static octahedral rotations, and thus it can induce surface polar states and pyroelectricity in a large class of otherwise nonpolar materials.

Oxide surfaces and interfaces exhibit intriguing properties such as two-dimensional electron gas, superconductivity [1, 2], charged domain walls [3], magnetism [4, 5] and multiferroicity [6]. Many oxide surfaces possess strong gradients of strain and octahedral rotations. Octahedral rotations (also called ***antiferrodistortions***) are the most common type of phase transitions involving lattice distortions in perovskite oxide systems [7]. Improper ferroelectricity induced by octahedral rotations is inherent in a number of oxides such as $YMnO_3$ [8], $Ca_3Mn_2O_7$ [9], $CaTiO_3$ [10, 11] and their interfaces [12]. Hereafter, we call the phenomena related to octahedral rotations as "roto" effects. Our primary interest is rotostriction induced by stress and strain field gradients at oxide interfaces through a quadratic coupling between octahedral rotations and strains.

It has been shown that strain and stress gradients [13, 14, 15, 16, 17] can induce polarization near the surfaces and interfaces via the flexoelectric effect [18, 19, 20]. Note that all materials are flexoelectrics [21], and all materials with static rotations (such as oxygen octahedra rotations) possess rotostriction. The joint action of the flexoelectric effect and rotostriction can thus lead to a ferroelectric polarization at an interface across which the octahedral rotation varies. Therefore, *every* antiferrodistortive boundary, twin wall, interface and surface can, in principle, possesses the roto-flexo effect. Since most functional oxide systems involve natural or artificial interfaces and surfaces, roto-flexo effects are quite general.

Experimental results show that surface influence systematically changes oxygen octahedral rotation behaviour [22, 23] (structural transitions in surface layers). Coexistence of antiferrodistortive and ferroelectric distortions was calculated from DFT at perovskite surfaces, such as $PbTiO_3$ (001) surface [24], while it is absent in $PbTiO_3$ bulk. In particular, DFT reconstruction of the $PbTiO_3$ (001) surface [25] revealed a single layer of antiferrodistortive structure with oxygen cages counter-rotated by 10 degree about the titanium ions. Antiferrodistortive reconstruction of the out-of-plane component of octahedral rotation was reached [24] at the PbO-terminated (001) surface and then observed with *x*-ray scattering [26]. DFT shows that tensile strain enhances the ferroelectric distortion and suppresses the antiferrodistortive rotation in the vicinity of $PbTiO_3$ (001) surface, while the opposite effect is caused by compressive strain [27].

Recently, we have theoretically predicted that a combination of flexoelectric effect and rotostriction at oxide interfaces can generate large improper ferroelectricity and pyroelectricity at antiferrodistortive boundaries and elastic twins in $SrTiO_3$ below 105K [28]. In this Letter we report that a polar state and pyroelectricity are induced by flexo-roto fields in the vicinity of ferroelastic $SrTiO_3$ surface even without any elastic domains.

Below we analyze the free energy functional corresponding to LGD expansion on the polar and structural order parameter components in the presence of ferroelastic surface. In the parent high temperature phase above the structural phase transition, the free energy density has the form:

$$G = \int_V d^3r\, g_b(\mathbf{r}) + \int_S d^3r\, g_S(\mathbf{r}) \tag{1}$$

in which the bulk ($g_b$) energy density is [28]:

$$g_b = g_P + g_\Phi + g_{grad} + g_{el} + g_{str} + g_{P-\Phi} + g_{flexo}, \tag{2a}$$

where polarization-dependent energy $g_P = a_i(T)P_i^2 + a_{ij}^u P_i^2 P_j^2 - P_i\left(\dfrac{E_i^d}{2} + E_i^{ext}\right)$, structural order parameter dependent energy $g_\Phi = b_i(T)\Phi_i^2 + b_{ij}^u \Phi_i^2 \Phi_j^2$, gradient energy $g_{grad} = \dfrac{g_{ijkl}}{2}\left(\dfrac{\partial P_i}{\partial x_j}\dfrac{\partial P_k}{\partial x_l}\right) + \dfrac{v_{ijkl}}{2}\left(\dfrac{\partial \Phi_i}{\partial x_j}\dfrac{\partial \Phi_k}{\partial x_l}\right)$, elastic strain energy $g_{el} = \dfrac{c_{ijkl}}{2} u_{ij} u_{kl}$, electro-striction and roto-striction energy $g_{str} = -q_{ijkl} u_{ij} P_k P_l - r_{ijkl} u_{ij} \Phi_k \Phi_l$, biquadratic coupling term $g_{P-\Phi} = -\eta_{ijkl}^u P_i P_j \Phi_k \Phi_l$ [29, 30, 31] and flexoelectric term $g_{flexo} = \dfrac{f_{ijkl}}{2}\left(\dfrac{\partial P_k}{\partial x_l} u_{ij} - P_k \dfrac{\partial u_{ij}}{\partial x_l}\right)$.

Summation is performed over all repeated indices.

$\Phi_i$ is the components ($i$=1 − 3) of the structural order parameter (OP), which is the vector corresponding to the spontaneous octahedral rotation angle around one of their fourfold symmetry axes in a structural phase [30, 32]. Note, that the rotation angle is proportional to the displacement (in pm) of an appropriate oxygen atom from its cubic position, as defined by Uwe and Sakudo [33]. $P_i$ is polarization vector, $u_{ij}(\mathbf{x})$ is the strain tensor ($i,j$ = 1 − 3). Gradients coefficients $g_{ij}$ and $v_{ij}$ are regarded positive for commensurate ferroics; $f_{ijkl}$ is the forth-rank tensor of flexoelectric coupling, $q_{ijkl}$ is the forth-rank electrostriction tensor, $r_{ijkl}$ is the rotostriction tensor, $c_{ijkl}$ is elastic stiffness. The flexoelectric effect tensor $f_{ijkl}$ and rotostriction tensor $r_{ijkl}$ have nonzero components in all phases and for any symmetry of the system. Tensors form for cubic $m3m$ symmetry is well-known; in particular $f_{12}$, $f_{11}$ and $f_{44}$ are nonzero for SrTiO$_3$ [34, 35]. Temperature dependence of coefficients $a_i$ and $b_i$ can be fitted with Barrett law, $a_i(T) = \alpha_T T_q^{(E)}\left(\coth(T_q^{(E)}/T) - \coth(T_q^{(E)}/T_0^{(E)})\right)$, $b_i(T) = \beta_T T_q^{(\Phi)}\left(\coth(T_q^{(\Phi)}/T) - \coth(T_q^{(\Phi)}/T_S)\right)$.

Surface ($g_S$) energy density is:

$$g_S = a_i^S P_i^2 + b_i^S \Phi_i^2 \qquad (2b)$$

Surface energy coefficients $a_i^S$ and $b_i^S$ ($i=1-3$) are regarded positive and weakly temperature dependent. Note that the values of $b_i^S$ could essentially influence near surface behaviour of the structural OP. For instance, the most likely case $b_3^S \ll b_{1,2}^S$ favors the octahedral rotations around the axis normal to the surface (as it was predicted by *ab initio* calculations for PbTiO$_3$ [24, 25, 27]).

Euler-Lagrange and elastic equations of state are obtained from the minimization of the free energy

$$\frac{\partial F_b}{\partial \Phi_i} = 0, \quad \frac{\partial F_b}{\partial P_i} = 0, \quad \frac{\partial F_b}{\partial u_{ij}} = \sigma_{ij}. \qquad (3)$$

where $\sigma_{ij}(\mathbf{x})$ is the stress tensor that satisfy mechanical equilibrium equation $\partial \sigma_{ij}(\mathbf{x})/\partial x_j = 0$. Equations (3) should be supplemented with the boundary conditions at $x_3 = 0$ for the OP vector:

$$\left( 2 b_i^S \Phi_i - v_{i3kl} \frac{\partial \Phi_k}{\partial x_l} \right)\bigg|_{x_3=0} = 0, \qquad (4)$$

polarization and stress, the latter two are modified by flexoelectric effect [19, 36]:

$$\left( 2 a_i^S P_i - g_{i3kl} \frac{\partial P_k}{\partial x_l} + \frac{f_{jki3}}{2} u_{jk} \right)\bigg|_{x_3=0} = 0, \qquad (5)$$

$$\left( \sigma_{3i} - \frac{f_{j3i3}}{2} \frac{\partial P_j}{\partial x_3} + \frac{f_{j3il}}{2} \frac{\partial P_j}{\partial x_l} \right)\bigg|_{x_3=0} = 0. \qquad (6)$$

Without flexoelectric effect the stress components $\sigma_{3i}(\mathbf{x}) = 0$ at the mechanically free surface $x_3 = 0$. Compatibility relations should be valid everywhere.

Hereafter we chose tetragonal SrTiO$_3$ ($T < 105$ K, space group I4/mcm) for numerical simulations, since all necessary parameters including gradient coefficients and flexoelectric tensor are known for the material (see **Table 1, Suppl. Mat**). Unfortunately exact values of gradient coefficients and flexoelectric tensor are unknown for other ferroelastics like CaTiO$_3$ or EuTiO$_3$, but the extension of the obtained results will be valid qualitatively for them, making the flexo-roto field induced polar states at surfaces and interfaces a general phenomenon in nature.

Now let us calculate the depth of the induced polarization penetration from the surface $x_3 \equiv z = 0$. For the case when 4-fold axis is *parallel* to the mono-domain SrTiO$_3$ surface, the *most thermodynamically preferable* situation is: two z-dependent components of OP vector, in-plane $\Phi_\parallel(z)$ and out-of plane $\Phi_\perp(z)$, and z-dependent in-plane polarization $P_\parallel(z)$ that does not cause any depolarization field ($E_\parallel^d = 0$, see the sketch of the problem geometry in **Fig. 1a**). Also one may consider out-of-plane polarization $P_\perp(z)$, but without enough concentration of free carriers its value is strongly affected by the depolarization field $-P_\perp(z)/\varepsilon_0\varepsilon_b$. We calculated numerically that $P_\parallel(z)$ values are at least $10^3$ times higher than $P_\perp(z)$ values without screening by free carriers.

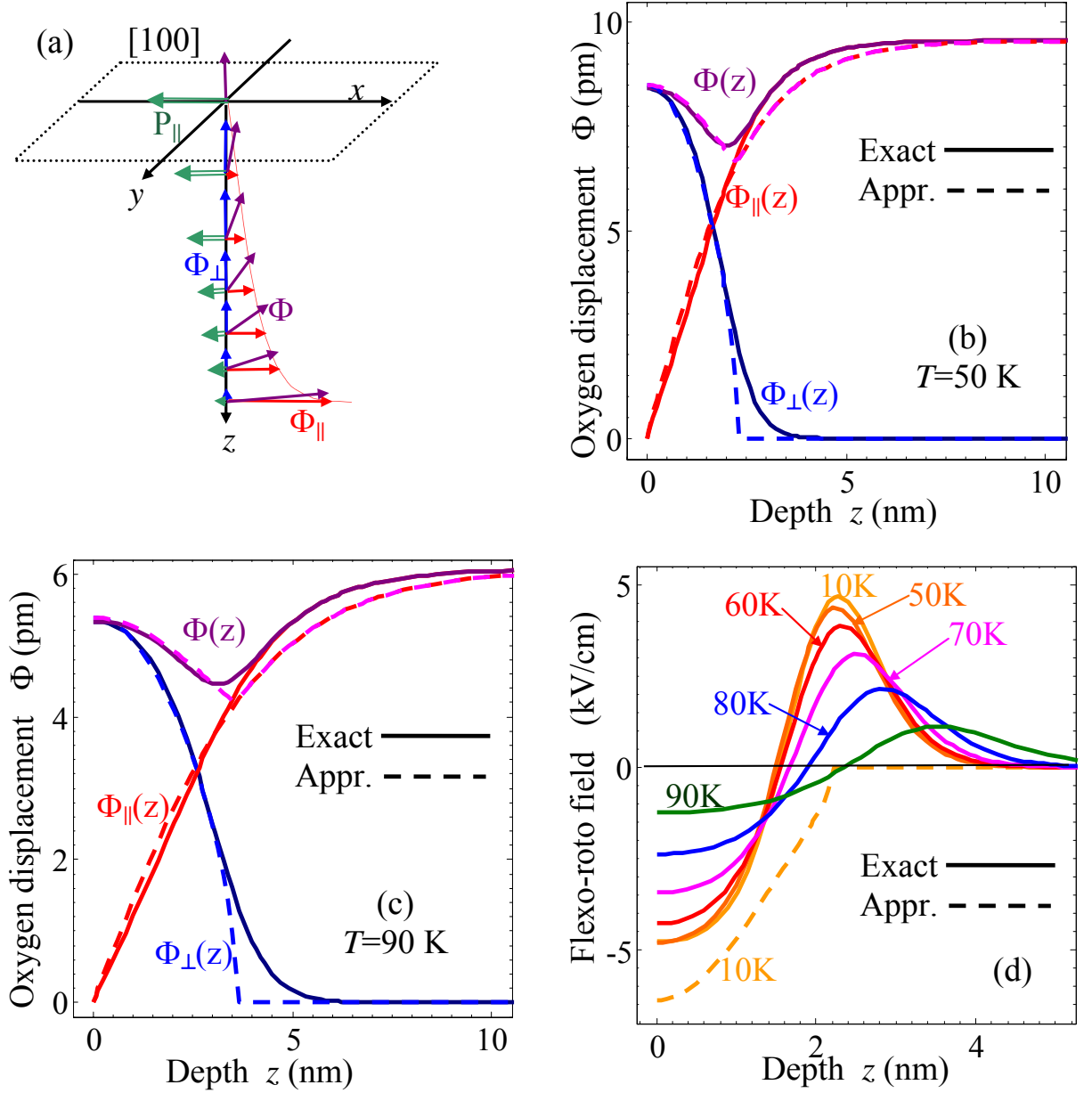

**Figure 1**. (a) Sketch of the problem geometry in the vicinity of SrTiO$_3$ [100] cut. 4-fold axis is parallel to the surface. (b, c) Depth z-profile of the structural OP components $\Phi_\perp(z)$, $\Phi_\parallel(z)$ and absolute value $\Phi(z)=\sqrt{\Phi_\perp^2(z)+\Phi_\parallel^2(z)}$ (labels near the curves) calculated numerically from *coupled* Eqs.(3) (solid curves) and analytically from *decoupled* Eqs.(8) (dashed curves) at temperature T= 50 K (b) and 90 K (c), SrTiO$_3$ parameters listed in the **Table 1, Suppl. Mat**,

extrapolation length $\lambda_\parallel = 0$ is defined after Eq.(8b). (d) Flexo-roto field $E_{FR}^B$ calculated at different temperatures 10, 50, 60, 70, 80 and 90 K (numbers near the curves).

For the case when 4-fold axis is *perpendicular* to the mono-domain SrTiO$_3$ surface, the OP becomes normal to the surface (i.e. out-of-plane) in the bulk of the sample. The appearance of in-plane OP components is not likely in this case (see comments after Eq.(2b)). As a result, only out-of-plane components of polarization $P_\perp(z)$ can be induced. The latter is strongly diminished by the depolarization field. Thus we do not consider the case here, especially because the length scale of $P_\perp(z)$ distribution is of order of lattice constant.

We numerically solve *coupled* system (3) when 4-fold axis is *parallel* to the mono-domain SrTiO$_3$ surface. Results are shown in **Figs. 1, 2,** and **3.** Our numerical simulations performed for coupled equations (3) with boundary conditions (4)-(6) demonstrate that polarization weakly affects structural OP. The fact makes it possible to *decouple* the polarization vector in the system (3) that reduces to the form (S.2), Suppl. Mat. Boundary conditions (6) acquire the form $\sigma_{3i}(\mathbf{x}) = 0$. The solution for strain and stresses has the form (S.3)-(S.4). Decoupling gives us the possibility to look for approximate analytical expression for OP and polarization. For the considered geometry, decoupled equations for OP components have the form:

$$2b_\perp \Phi_\perp + 4b_{11}^\perp \Phi_\perp^3 + 2b_c \Phi_\perp \Phi_\parallel^2 - v_{11} \frac{\partial^2 \Phi_\perp}{\partial z^2} = 0, \quad (7a)$$

$$2b_\parallel \Phi_\parallel + 4b_{11}^\parallel \Phi_\parallel^3 + 2b_c \Phi_\parallel \Phi_\perp^2 - v_{44} \frac{\partial^2 \Phi_\parallel}{\partial z^2} = 0, \quad (7b)$$

where the following designations are introduced: $b_\perp = b_1 - \left( b_{12}^u - b_{12}^\sigma - \frac{r_{11} r_{12}}{c_{11}} + \frac{r_{44}^2}{2c_{44}} \right) \Phi_B^2$,

$b_{11}^\perp = b_{11}^u - \frac{r_{11}^2}{2c_{11}}$, $b_\parallel = b_1 - 2\left( b_{11}^u - b_{11}^\sigma - \frac{r_{12}^2}{2c_{11}} \right) \Phi_B^2$, $b_{11}^\parallel = b_{11}^u - \frac{r_{12}^2}{2c_{11}}$, $b_c = b_{12}^u - \frac{r_{11} r_{12}}{c_{11}} - \frac{r_{44}^2}{2c_{44}}$. Also

we used expansion coefficients at given stress, $b_{ijkl}^\sigma = b_{ijkl}^u - r_{msji} s_{mspq} r_{pqkl}/2$, here $s_{mnij}$ is the elastic compliances tensor. The bulk value of OP is $\Phi_B(T) = \sqrt{-b_1(T)/2b_{11}^\sigma}$. Since the condition $v_{11} \ll v_{44}$ is typically valid due to the fact of strong coupling of octahedron rotations in the layer, perpendicular to rotation axis, and weak coupling between such layers [37], the second derivative can be neglected in Eq.(7a) [30]. So under the condition $b_\perp + b_c \Phi_\parallel^2(z) < 0$, approximate solution acquires the form:

$$\Phi_\perp(z) \approx \sqrt{-\frac{1}{2b_{11}^\perp}\left(b_\perp + b_c \Phi_\|^2(z)\right)} \qquad (8a)$$

while $\Phi_\perp(z) \equiv 0$ and $b_\perp + b_c \Phi_\|^2(z) > 0$. Solution (8a) is not consistent with the boundary conditions (4a) for $\Phi_\perp(0)$ in general case, but our numerical simulations proved that the influence of the boundary condition becomes negligible even at very small distances from the surface. This happens because corresponding length scale $L_\perp(T) = \sqrt{v_{11}/|b_\perp(T)|}$ is smaller that the lattice constant making approximation (8a) self-consistent. Solution for the OP component $\Phi_\|(z)$ was derived by direct variation method as:

$$\Phi_\|(z) \approx \Phi_B \cdot \tanh\left(\frac{z-z_0}{L_\Phi}\right) \approx \Phi_B\left(1 - \frac{1}{1+\sqrt{2}\lambda_\|/L_\Phi}\exp\left(-\frac{\sqrt{2}z}{L_\Phi}\right)\right), \qquad (8b)$$

Correlation length is introduced as $L_\Phi(T) = \sqrt{2v_{44}/|b_\|(T)|}$; the extrapolation lengths for "in-plane" component of OP is introduced as $\lambda_\| = v_{44}/b_1^S$. The length is determined by the surface energy (2b) coefficient $b_1^S \geq 0$. Hereinafter we regard the extrapolation length to be not negative, otherwise higher positively defined terms should be included in the surface free energy (1c). Note, that the case $\lambda_\| = 0$ (i.e. $z_0 = 0$) corresponds to maximal possible amplitude $\Phi_\perp(0) = \max$ and minimal $\Phi_\|(0) = 0$ [as shown in **Fig. 1b**]. At arbitrary $\lambda_\|$ constant $z_0$ found from the boundary condition (4a) is $z_0 = L_\Phi \times \text{actanh}\left((\lambda_\|/L_\Phi)\left(0.5 + \sqrt{0.25 + (\lambda_\|/L_\Phi)^2}\right)^{-1}\right)$. The gradient region under the surface has the maximal depth exactly for the case of $z_0 = 0$. In the general case, the characteristic depth of the gradient region is about several $L_\Phi$.

Numerical simulations proved that the approximate analytical expressions (8) relatively accurately reproduce the OP distribution calculated numerically from Eq.(3) and their gradients in the near-surface region (see **Fig. 1b-c**).

Using the elastic solution (S.3)-(S.4) and decoupling approximation, we simplify equation for polarization $\partial F_b/\partial P_i = 0$ as:

$$\alpha_1(z)P_\| + \left(4a_{11}^u - 2\frac{q_{12}^2}{c_{11}}\right)P_\|^3 - \left(g_{44} - \frac{f_{44}^2}{c_{44}}\right)\frac{\partial^2 P_\|}{\partial z^2} = E_{FR}^B(z), \qquad (9a)$$

$$\left.\left(P_\| - \lambda_P \frac{\partial P_\|}{\partial z} + P_{FR}^S\right)\right|_{z=0} = 0. \qquad (9b)$$

So-called polarization extrapolation length is introduced as $\lambda_P = g_{44}/2a_1^S$, whose geometrical sense is described in Ref.[38]. The length is determined by the surface energy (2b) coefficient $a_1^S$ that depends on the surface state and is poorly known for ferroelectrics [39]. Since $\lambda_P$ is unknown for SrTiO$_3$, we vary it in the physically realistic range of 1 – 100 nm.

It follows from Eqs.(9) that there are several sources of the polarization appearance in the vicinity of surface. The first source is the inhomogeneity in the right-hand-side of Eq.(9a): electric field $E_{FR}^B(z) = \frac{r_{44} f_{44}}{c_{44}} \frac{\partial(\Phi_\| \Phi_\perp)}{\partial z}$, which strength is proportional to the convolution of the flexoelectric and rotostriction tensors with OP gradient, further regarded as *gradient flexo-roto field*. Depth profile of $E_{FR}(z)$ is shown in **Fig. 1d.** The second source is the inhomogeneity in the boundary conditions (9b), $P_{FR}^S = \frac{r_{44} f_{44}}{4 c_{44} a_1^S} \Phi_\|(0) \Phi_\perp(0)$, whose strength is also proportional to the convolution of the flexoelectric and rotostriction tensors with OP, further regarded as *built-in surface flexo-roto polarization* Both these sources induce *improper spontaneous polarization*. Note, that $P_{FR}^S = 0$ for the case $\lambda_\| = 0$, since $\Phi_\|(0) = 0$.

The condition $\alpha_1(z) < 0$, that are valid near the surface at low temperatures, can lead to the *roto-induced ferroelectric polarization* appearance under negligibly small depolarization field. Estimations made for SrTiO$_3$ parameters prove that coefficient

$$\alpha_1(z) = 2\left(a_1 - \left(\eta_{11}^u + \frac{q_{12} r_{12}}{c_{11}}\right)\Phi_\|^2(z) - \left(\eta_{11}^\sigma - \eta_{11}^u - \frac{q_{12} r_{12}}{c_{11}}\right)\Phi_B^2 - \left(\eta_{12}^u + \frac{q_{12} r_{11}}{c_{11}}\right)\Phi_\perp^2(z)\right) \quad (10)$$

is positive in the single-domain bulk material at temperature $T < T_S$, where $\Phi_\|(z) \approx \Phi_B$ (otherwise a bulk material should be ferroelectric). Here we re-introduced the biquadratic coupling coefficient $\eta_{ijkl}^\sigma = \eta_{ijkl}^u + q_{msji} s_{mspq} r_{pqkl}$. However $\alpha_1(z)$ is strongly coordinate-dependent near the surface $z = 0$ as shown in **Fig. 2a.** For SrTiO$_3$ parameters $\alpha_1 > 0$ at temperatures $T > 50$ K and it becomes negative due to the biquadratic coupling at $T < 50$ K.

Nonlinearity and gradients terms can be omitted in the Eq.(9a) in the region where $\alpha_1 > 0$, leading to the simple approximate expression for polarization distribution:

$$P_\|(z) \sim \frac{f_{44} r_{44}}{\alpha_1(z) c_{44}} \frac{\partial(\Phi_\| \Phi_\perp)}{\partial z} \quad (11)$$

Spontaneous polarization (11) is "incipient" as induced by the flexo-roto coupling, and thus no ferroelectric hysteresis exists at temperatures $T > 50$ K. True ferroelectricity appears and hysteresis loop opens for $\alpha_1 < 0$, that is possible at $0 \leq z \leq 6$ nm and $T > 50$ K.

From **Fig. 2b** we can conclude that the flexo-roto fields do induce polar state under the surface at distances $z \leq 2L_\Phi$ in ferroelastics. Note, that $L_\Phi \sim 3$ nm for SrTiO$_3$ at T<90 K [28, 30] determines the nanometer scale of the surface polar state. So, the typical thickness of polar state is about 7 lattice constants, making continuum theory results at least semi-quantitatively valid. Polarization appears at temperatures lower than $T_S$ and it increases as the temperature decreases (compare different curves in **Fig. 2b**). Surface polarization and maximal values increase as the extrapolation length $\lambda_P$ increases (compare different curves in **Fig. 2d**). For small $\lambda_P$ polarization may change its sign in the interface region (see **Fig. 2c**). It is seen that spontaneous polarization can reach noticeable values $\sim 1 - 10$ μC/cm$^2$ in the gradient region $z \leq 2L_\Phi$ at temperatures lower than 60 K.

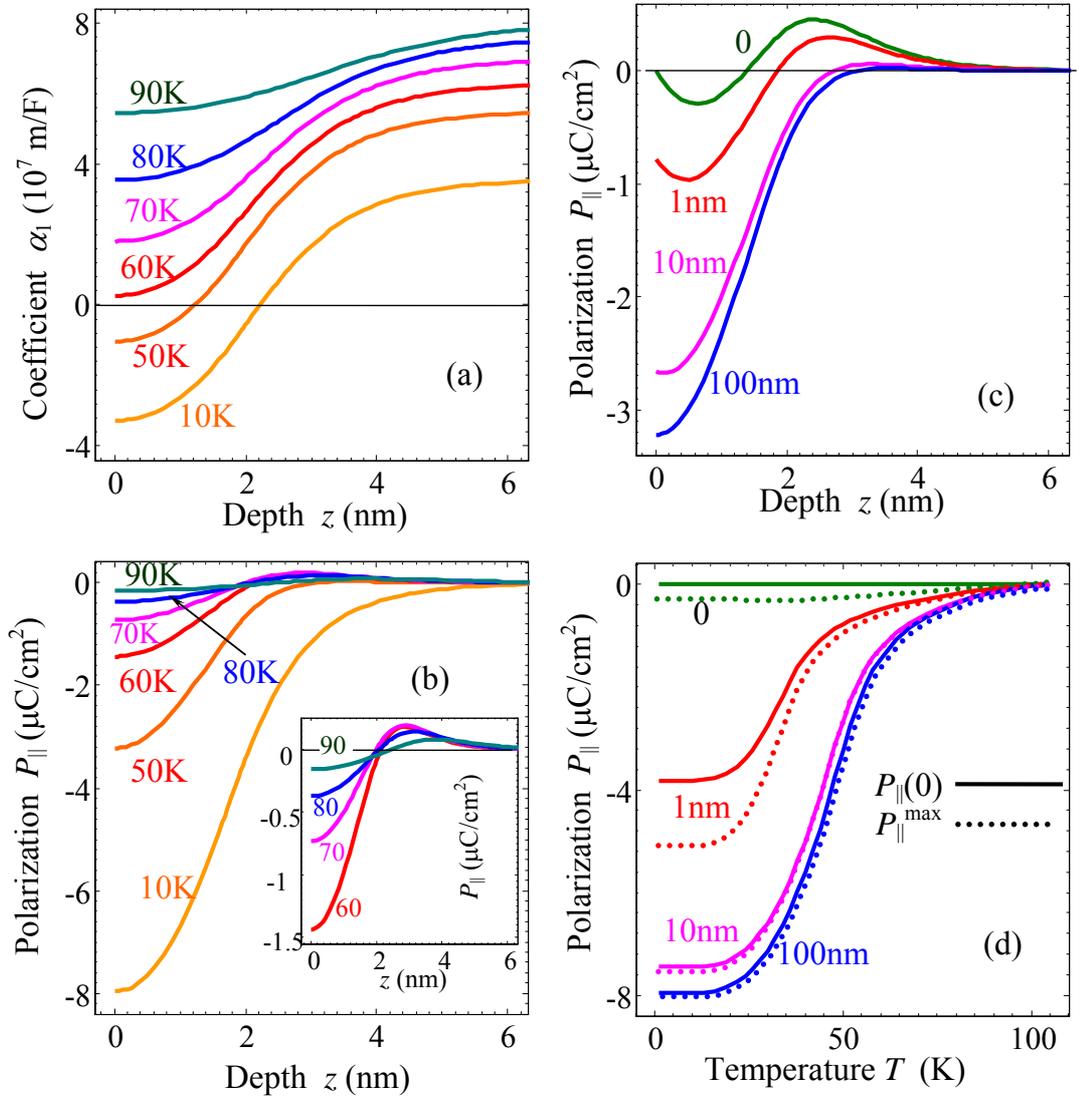

**Figure 2**. (a) Coefficient $\alpha_1(z)$ and (b) spontaneous polarization $P_\parallel(z)$ vs. the depth z from the surface calculated numerically from coupled Eqs.(3) at different temperatures 10, 50, 60, 70, 80

and 90 K (numbers near the curves) for polarization extrapolation length $\lambda_P = 0$. (c) Polarization $P_\|(z)$ vs. the depth z calculated for different length $\lambda_P = 0$, 1 nm, 10 nm, 100 nm (figures near the curves) and temperature 50 K. (d) Surface polarization $P_\|(0)$ (solid curves) and polarization maximal value $P_\|^{max}(z)$ (dotted curves) vs. temperature calculated for $\lambda_P = 0$, 1 nm, 10 nm, 100 nm (figures near the curves). Material parameters of SrTiO₃ are listed in the **Table 1, Suppl. Mat.,** structural OP extrapolation length $\lambda_\| = 0$.

Temperature dependence of polarization $\langle P_\| \rangle$ averaged over the polar layer thickness $w$=5 nm is shown in **Fig. 3a.** Temperature dependence of pyroelectric coefficient $\langle \Pi_\| \rangle$ averaged over the polar layer thickness $w$=5 nm is shown in **Fig. 3b.** We calculated noticeable pyroelectric coefficient $\langle \Pi_\| \rangle \sim 2 \times 10^{-3}$ C/m²K. The values are well above detectable limits of pyroelectric coefficient, which are about $10^{-6}$ C/m²K [40]. Thus either planar electrode setup or PyroSPM [41] supplied with sharp tips of sizes 5-10 nm could reliably detect local lateral pyroelectric response of the ferroelastic surface.

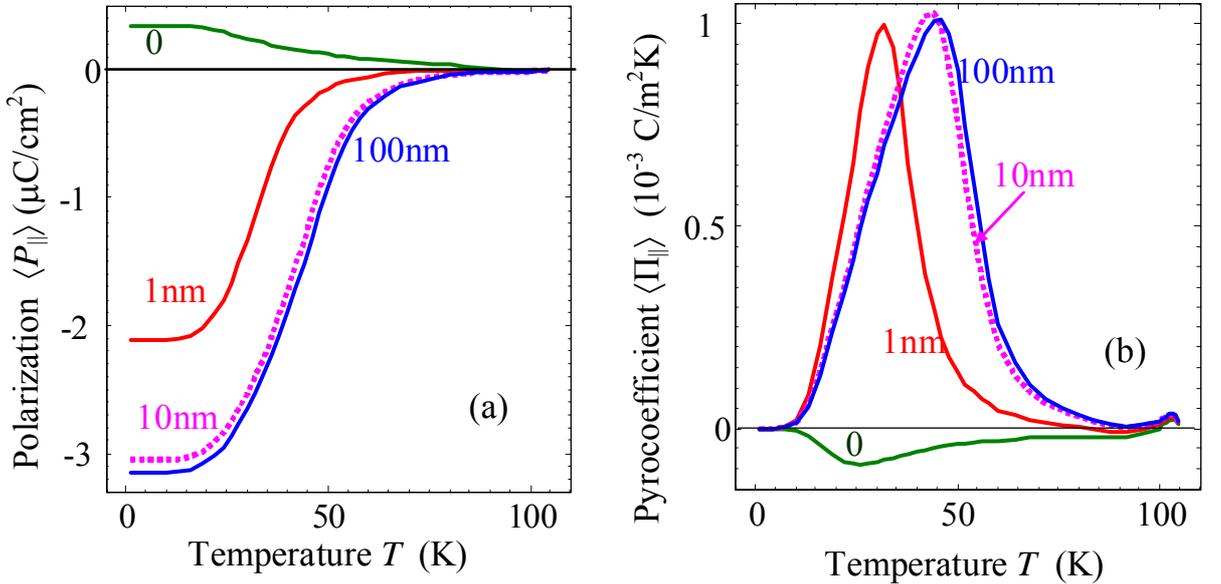

**Figure 3**. Temperature dependence of average polarization $\langle P_\| \rangle$ (a) and pyroelectric coefficient $\langle \Pi_\| \rangle$ (b) calculated from coupled Eqs.(3) for different length $\lambda_P = 0$, 1 nm, 10 nm, 100 nm (figures near the curves). Material parameters of SrTiO₃ are listed in the **Table 1, Suppl. Mat,** extrapolation length $\lambda_\| = 0$.

**To summarize,** we report a new source of electric field, which we name gradient *flexo-roto field*, induces a significant improper spontaneous polarization and pyroelectricity in the vicinity of surfaces and interfaces of otherwise non-ferroelectric ferroelastics such as $SrTiO_3$, and by extension in $CaTiO_3$, $EuTiO_3$, and in antiferroelectrics like $PbZrO_3$. In $SrTiO_3$ flexo-roto effect leads to a large spontaneous polarization ($\sim 1 - 5$ $\mu C/cm^2$) and pyroelectric coefficient ($\sim 10^{-3}$ $C/m^2 K$). The strength of the gradient flexo-roto field is proportional to the convolution of the flexoelectric and rotostriction tensors with the gradients of octahedral rotations, which are structural order parameters. The strength of the surface flexo-roto polarization is proportional to the convolution of the flexoelectric and rotostriction tensors with octahedral rotations on the surface. Flexo-roto effects should exist at surfaces in all structures with static rotations, which are abundant in nature, it allows for contribution into polar interfaces in a large class of nonpolar materials.

Note, that there are other possible reasons for polar surface states in nonpolar materials such as $SrTiO_3$: space charge due to defect chemistry and band gap differences between surfaces and bulk [42], surface reconstruction and atom clustering [43], surface piezoelectricity [44, 45] and strained polar regions that extends into the bulk at a distance much larger than a few nanometers [46]. In accordance with these and other studies, combined rotostriction and flexoelectricity cannot not be the sole contribution to the polar surface stats in ferroelastics. However the conclusion in this Letter is that the surfaces of all ferroelastics with octahedral tilts should be intrinsically polar in the low temperature octahedrally tilted phase.

Authors gratefully acknowledge multiple discussions with S.L. Bravina (NASU) and Zheng Gai (ORNL). S.V.K. acknowledges FWP. V.G. and L.Q.C. would like to acknowledge funding from the National Science Foundation grant numbers DMR-0908718 and DMR-0820404.

**Supplementary Materials**

Euler-Lagrange equations of state (3) obtained from the minimization of the free energy (1) are

$$2 b_i \Phi_i + 4 b_{ij}^u \Phi_j^2 \Phi_i - v_{ijkl} \frac{\partial^2 \Phi_k}{\partial x_j \partial x_l} - 2 r_{mjki} u_{mj} \Phi_k - 2\eta_{klij}^u P_k P_l \Phi_j = 0, \quad (S.1a)$$

$$c_{ijkl} u_{kl} - r_{ijkl} \Phi_k \Phi_l + f_{ijkl} \frac{\partial P_k}{\partial x_l} - q_{ijkl} P_k P_l = \sigma_{ij}, \quad (S.1b)$$

$$2 a_i P_i + 4 a_{ijkl}^u P_j P_k P_l - g_{ijkl} \frac{\partial^2 P_k}{\partial x_j \partial x_l} - 2 q_{mjki} u_{mj} P_k - f_{mnil} \frac{\partial u_{mn}}{\partial x_l} - 2\eta_{ijkl}^u P_j \Phi_k \Phi_l = E_i^d. \quad (S.1c)$$

Decoupling on polarization vector in Eqs.(S.1) leads to:

$$2 b_i \Phi_i + 4 b_{ij}^u \Phi_j^2 \Phi_i - v_{ijkl} \frac{\partial^2 \Phi_k}{\partial x_j \partial x_l} - 2 r_{mjki} u_{mj} \Phi_k = 0, \quad (S.2a)$$

$$2 a_i P_i - 2\eta_{ijkl}^u P_j \Phi_k \Phi_l + 4 a_{ijkl}^u P_j P_k P_l - g_{ijkl} \frac{\partial^2 P_k}{\partial x_j \partial x_l} - 2 q_{mjki} u_{mj} P_k - f_{mnil} \frac{\partial u_{mn}}{\partial x_l} = E_i^d, \quad (S.2b)$$

$$c_{ijkl} u_{kl} = \sigma_{ij} + r_{ijkl} \Phi_k \Phi_l. \quad (S.2c)$$

For considered geometry (see **Fig. 1a**) elastic solution for strain tensor in Voigt notations (11=1, 22=2, 33=3, 23=4, 13=5, 12=6) has the form:

$$u_1 = \frac{(c_{11} + c_{12}) r_{11} - 2 c_{12} r_{12}}{(c_{11} - c_{12})(c_{11} + 2c_{12})} \Phi_B^2, \quad u_2 = \frac{c_{11} r_{12} - c_{12} r_{11}}{(c_{11} - c_{12})(c_{11} + 2c_{12})} \Phi_B^2, \quad (S.3a)$$

$$u_3 = \frac{q_{12} P_1^2 + r_{11} \Phi_3^2 + r_{12} \Phi_1^2}{c_{11}} - \frac{c_{12}}{c_{11}} \left( \frac{c_{11} r_{11} + (c_{11} - 2c_{12}) r_{12}}{(c_{11} - c_{12})(c_{11} + 2c_{12})} \right) \Phi_B^2, \quad (S.3b)$$

$$u_4 = 0, \quad u_5 = -\frac{f_{44}}{c_{44}} \frac{\partial P_1}{\partial x_3} + \frac{r_{44}}{c_{44}} \Phi_1 \Phi_3, \quad u_6 = 0. \quad (S.3c)$$

Stress tensor components in Voigt notations has the form

$$\sigma_1 = \left( r_{11} - r_{12} \frac{c_{12}}{c_{11}} \right)(\Phi_B^2 - \Phi_1^2) + \left( r_{11} \frac{c_{12}}{c_{11}} - r_{12} \right) \Phi_3^2 + \left( q_{12} \frac{c_{12}}{c_{11}} - q_{11} \right) P_1^2, \quad (S.4a)$$

$$\sigma_2 = r_{12} \left( 1 - \frac{c_{12}}{c_{11}} \right)(\Phi_B^2 - \Phi_1^2) + \left( r_{11} \frac{c_{12}}{c_{11}} - r_{12} \right) \Phi_3^2 + q_{12} \left( \frac{c_{12}}{c_{11}} - 1 \right) P_1^2, \quad (S.4b)$$

$$\sigma_3 = \sigma_4 = \sigma_5 = \sigma_6 = 0 \quad (S.4c)$$

Substituting Eqs.(S.3, 4) in Eq.(S.1) we get coupled system of equations for structural order parameters and polarization. Polarization satisfies the following equation:

$$2\left( a_1 - \left( \eta_{11}^u + \frac{q_{12} r_{12}}{c_{11}} \right) \Phi_1^2 - \left( \eta_{11}^\sigma - \eta_{11}^u - \frac{q_{12} r_{12}}{c_{11}} \right) \Phi_B^2 - \left( \eta_{12}^u + \frac{q_{12} r_{11}}{c_{11}} \right) \Phi_3^2 \right) P_1$$

$$+ \left( 4 a_{11}^u - 2 \frac{q_{12}^2}{c_{11}} \right) P_1^3 - \left( g_{44} - \frac{f_{44}^2}{c_{44}} \right) \frac{\partial^2 P_1}{\partial x_3^2} = \frac{r_{44} f_{44}}{c_{44}} \frac{\partial(\Phi_1 \Phi_3)}{\partial x_3} \quad (S.5)$$

Here we introduced the expansion coefficients at given stress. Namely, for coupling coefficients

$$\eta_{11}^{\sigma} = \eta_{11}^{u} + \frac{2(q_{11}-q_{12})(r_{11}-r_{12})}{3(c_{11}-c_{12})} + \frac{(q_{11}+2q_{12})(r_{11}+2r_{12})}{3(c_{11}+2c_{12})} \quad \text{(S.6a)}$$

$$\eta_{12}^{\sigma} = \eta_{12}^{u} - \frac{(q_{11}-q_{12})(r_{11}-r_{12})}{3(c_{11}-c_{12})} + \frac{(q_{11}+2q_{12})(r_{11}+2r_{12})}{3(c_{11}+2c_{12})} \quad \text{(S.6b)}$$

$$\eta_{44}^{\sigma} = \eta_{44}^{u} + \frac{q_{44}r_{44}}{c_{44}} \quad \text{(S.6c)}$$

And for nonlinearity coefficients

$$b_{11}^{\sigma} = b_{11}^{u} - \frac{1(r_{11}-r_{12})^2}{3(c_{11}-c_{12})} - \frac{(r_{11}+2r_{12})^2}{6(c_{11}+2c_{12})}, \quad \text{(S.7a)}$$

$$b_{12}^{\sigma} = b_{12}^{u} + \frac{1(r_{11}-r_{12})^2}{3(c_{11}-c_{12})} - \frac{(r_{11}+2r_{12})^2}{3(c_{11}+2c_{12})} - \frac{r_{44}^2}{2c_{44}} \quad \text{(S.7b)}$$

$$a_{11}^{\sigma} = a_{11}^{u} - \frac{1(q_{11}-q_{12})^2}{3(c_{11}-c_{12})} - \frac{(q_{11}+2q_{12})^2}{6(c_{11}+2c_{12})}, \quad \text{(S.8a)}$$

$$a_{12}^{\sigma} = a_{12}^{u} + \frac{1(q_{11}-q_{12})^2}{3(c_{11}-c_{12})} - \frac{(q_{11}+2q_{12})^2}{3(c_{11}+2c_{12})} - \frac{q_{44}^2}{2c_{44}} \quad \text{(S.8b)}$$

**Table 1**. SrTiO$_3$ material parameters and LGD free energy (1)

| Parameter | SI units | Value | Source and notes |
|---|---|---|---|
| $\varepsilon_b$ | dimensionless | 43 | [a, b] |
| $\alpha_T$ | $10^6 \times$m/(F K) | 0.75 | [c, d] |
| $T_0^{(E)}$ | K | 30 | ibidem |
| $T_q^{(E)}$ | K | 54 | ibidem |
| $a_{ij}$ | $10^9 \times$m$^5$/(C$^2$F) | $a_{11}^u$=2.025, $a_{12}^u$=1.215, $a_{11}^\sigma$=1.724, $a_{12}^\sigma$=1.396 | ibidem<br>calculated from $a_{ij}^u$ |
| $q_{ij}$ | $10^{10}\times$ m/F | $q_{11}$=1.251, $q_{12}$= –0.108, $q_{44}$=0.243 | [c] |
| $g_{ijkl}$ | $10^{-11}\times$V·m$^3$/C | $g_{11}=g_{44}$=1, $g_{12}$=0.5 | Estimation based on Ref. [e] |
| $\beta_T$ | $10^{26}\times$J/(m$^5$ K) | 9.1 | [c] |
| $T_S$ | K | 105 | [c] |
| $T_q^{(\Phi)}$ | K | 145 | [c] |
| $b_{ij}$ | $10^{50}\times$J/m$^7$ | $b_{11}^u$=1.94, $b_{12}^u$=3.96, $b_{11}^\sigma$=1.69, $b_{12}^\sigma$=3.88 | [c]<br>calculated from $b_{ij}^u$ |
| $r_{ij}$ | $10^{30}\times$J/(m$^5$) | $r_{11}$=1.3, $r_{12}$= –2.5, $r_{44}$=–2.3 | [c] |
| $\eta_{ijkl}$ | $10^{29}$ (F m)$^{-1}$ | $\eta_{11}^u$=–3.366, $\eta_{12}^u$= 0.135, $\eta_{44}^u$=6.3<br>$\eta_{11}^\sigma$=–2.095, $\eta_{12}^\sigma$= –0.849, $\eta_{44}^\sigma$=5.860 | [c]<br>calculated from $\eta_{ij}^u$ |
| $v_{ijkl}$ | $10^{10}\times$J/m$^3$ | $v_{11}$=0.28, $v_{12}$= –7.34, $v_{44}$=7.11 | [d, f] |
| $c_{ij}$ | $10^{11}\times$J/m$^3$ | $c_{11}$=3.36, $c_{12}$=1.07, $c_{44}$=1.27 | [c, d] |

| $s_{ij}$ | $10^{-12} \times m^3/J$ | $s_{11}$=3.52, $s_{12}$= −0.85, $s_{44}$=7.87 | calculated from $c_{ij}$ |
| $f_{ijkl}$ | V | $f_{11}^e$ = − 3.24 , $f_{12}^e$ = 1.44 , $f_{44}^e$ = 1.08 | Recalculated from Ref.[g] at given stress |

[a] G. Rupprecht and R.O. Bell, Phys. Rev. **135**, A748 (1964).

[b] G.A. Smolenskii, V.A. Bokov, V.A. Isupov, N.N Krainik, R.E. Pasynkov, A.I. Sokolov, *Ferroelectrics and Related Materials* (Gordon and Breach, New York, 1984). P. 421

[c] N. A. Pertsev, A. K. Tagantsev, and N. Setter, Phys. Rev. B **61**, R825 (2000).

[d] A.K. Tagantsev, E. Courtens and L. Arzel, Phys. Rev. B, **64**, 224107 (2001).

[e] J. Hlinka and P. Marton, Phys. Rev. B **74**, 104104 (2006).

[f] W. Cao and R. Barsch, Phys. Rev. B, **41**, 4334 (1990)

[g] P. Zubko, G. Catalan, A. Buckley, P.R. L. Welche, J. F. Scott. Phys. Rev. Lett. **99**, 167601 (2007).